\documentclass[12pt]{article}

\newcommand{\be}{\begin{equation}}
\newcommand{\ee}{\end{equation}}
\newcommand{\bi}[1]{\vspace{-3mm} \bibitem{#1}}

\usepackage{epsfig,amsmath,amssymb,graphics,graphicx} 

\begin{document}
%%%%%%%%%%%%%%%%%%%%%%%%%%%%%%%%%%%%%%%%%%%%%%%%%%%%%%%%%%%%%%%%%%%%%%%%%%%%%%%%%%%%%%%%%%%%%%%%%%%%
\begin{center}

{\bf \large Towards Fractional Gradient Elasticity} \\ 

\vskip 7mm
{\bf \large Vasily E. Tarasov} \\
\vskip 3mm

{\it Skobeltsyn Institute of Nuclear Physics,\\ 
Lomonosov Moscow State University, Moscow 119991, Russia} \\
{E-mail: tarasov@theory.sinp.msu.ru} \\

\vskip 7mm
{\bf \large Elias C. Aifantis}\footnote{Emeritus Professor of Engineering,
Michigan Tech, Houghton, MI 49931, USA\\
Distinguished Adjunct Professor of King Abdulaziz University,
Jeddah, 21589, SA} \\
\vskip 3mm

{\it Laboratory of Mechanics and Materials, \\ 
 Aristotle University of Thessaloniki, Thessaloniki 54006, Greece} \\
{E-mail: mom@mom.gen.auth.gr} \\

\begin{abstract} 
An extension of gradient elasticity through the inclusion of spatial derivatives 
of fractional order to describe power-law type of non-locality is discussed.
Two phenomenological possibilities are explored.
The first is based on the Caputo fractional derivatives in one-dimension.
The second involves the Riesz fractional derivative in three-dimensions.
Explicit solutions of the corresponding fractional differential equations
are obtained in both cases.
In the first case it is shown that stress equilibrium in a Caputo elastic bar
requires the existence of a non-zero internal body force to equilibrate it.
In the second case, it is shown that in a Riesz type gradient elastic continuum 
under the action of a point load, the displacement may or may not be singular
depending on the order of the fractional derivative assumed.
\end{abstract}

\end{center}

%%% 45.10.Hj Perturbation and fractional calculus methods
%%% 62.20.Dc Elasticity, elastic constants.
%%% 61.50.Ah Theory of crystal structure, crystal symmetry; calculations and modeling
%%% 81.40.Jj Elasticity and anelasticity, stress-strain relations

%%%%%%%%%%%%%%%%%%%%%%%%%%%%%%%%%%%%%%%%%%%%%%%%%%%%%%%%%%%%%%%%%%%%%%%%%%%%%%%%%%%%%%%%%%%%%%%%%%%%

\section{Introduction}

The use of fractional derivatives and integrals \cite{SKM,KST} 
allow us to investigate the behavior of materials processes 
and systems that are characterized by power-law non-locality, 
power-law long term memory and fractal properties.
Fractional calculus has emerged as a powerful tool to consider
a wide range of applications in mechanics 
and physics (e.g. \cite{CM} - \cite{IJMPB2013}).

Non-local effects in elasticity theory have been 
treated with two different approaches:
the gradient elasticity theory (weak non-locality) and 
the integral elasticity theory (strong non-locality).
The fractional calculus can then be used to establish 
a fractional generalization of non-local elasticity in two forms: 
the fractional gradient elasticity theory (weak non-locality) and 
the fractional integral elasticity theory (strong non-locality).

Some developments 
a framework and derivation of corresponding results
for the fractional integral elasticity have been made in \cite{Laz,CPZ-1,CCS-2}.
%%%as proposed by the second author 
This has not be done, however, for the case of gradient elasticity 
(for a recent review of the subject one may consult \cite{A2011,AA2011}). 
An extension of the phenomenological theory of gradient elasticity by including 
the Caputo and Riesz spatial derivatives of non-integer order is suggested in the present paper. 

%%%An extension of the phenomenological theory of gradient elasticity 
%%%by including spatial derivatives of non-integer order,
%%%and its realization with the Caputo and Riesz fractional derivatives is suggested
%%%in the present paper. 

%%% An extension of the phenomenological theory of gradient elasticity 
%%% by including spatial derivatives of non-integer orders has been suggested by the second author,
%%% and its realization with the Caputo and Riesz fractional derivatives has been performed 
%%% by the first author using the mathematical framework and corresponding results 
%%% developed in \cite{TarasovSpringer}.

In Section 2 a phenomenological fractional generalization of 
one-dimensional gradient  elasticity is discussed by employing
the Caputo derivative to include gradient effects 
in the constitutive equation for the stress.
The corresponding fractional differential equation for the displacement
is solved analytically and expressed in terms 
of the Mittag-Leffler functions.
Among other things it is shown that in order that 
the stress field be equilibrated, the material develops an
internal force that is added to the externally applied body force field.

In Section 3 a fractional generalization of gradient elasticity 
is discussed by employing the Riesz derivative 
(in particular, the fractional Laplacian in the Riesz form). 
Analytical solutions of the corresponding fractional differential equation 
are obtained for two cases: 
the sub-gradient and super-gradient elasticity
(in analogy to sub-diffusion and super-diffusion case)
for a continuum carrying a point load.
Asymptotic expressions are derived for the displacement field
near the point of application of the external load. They may or may not
be singular depending on the order of the fractional derivative used.

%%%%%%%%%%%%%%%%%%%%%%%%%%%%%%%%%%%%%%%%%%%%%%%%%%%%%%%%%%%%%%%%%%%%%%%%%%%%%%

\section{Fractional gradient elasticity based on Caputo derivative}

%%%%%%%%%%%%%%%%%%%%%%%%%%%%%%%%%%%%%%%%%%%%%%%%%%%%%%%%%%%%%%%%%%%%%%%%%%%%%%

In this section we suggest a fractional generalization of the gradient elasticity model
that includes the Caputo derivative of non-integer order.
For this one-dimensional model we derive a general solution of
the corresponding fractional differential equation for the displacement.
We demonstrate how to overcome the difficulties caused by 
the unusual properties of fractional derivatives.
An alternative fractional gradient elasticity model for a three-dimensional case
employing the Riesz fractional derivative
(in the form of a fractional Laplacian) is discussed in Section 3.

%%%%%%%%%%%%%%%%%%%%%%%%%%%%%%%%%%%%%%%%%%%%%%%%%%%%%%%%%%%%%%%%%%%%%%%%%%%%%%

\subsection{Fractional gradient elasticity equation}

Let us consider the constitutive relation for a one-dimensional
fractional gradient elasticity model 
that is based on the Caputo derivative in the form
\be 
\sigma(x) = E \, \varepsilon (x) \pm l^2_{\beta} \, E \ ^CD^{\beta}_{a+} \varepsilon (x) ,
\ee
where $\sigma(x)$ is the stress, $\varepsilon (x)$ is the strain, 
with the space variable $x$ and the scale parameter $l^2_{\beta}$ being dimensionless. 
The symbol $^CD^{\beta}_{a+}$ is the Caputo derivative of order $\beta$ with $n-1<\beta<n$. 
The $\pm$ sign is kept for generality as various previous non-fractional 
gradient elasticity models employ either sign 
(for a comprehensive of non-fractional gradient elasticity models the readers may consult \cite{AA2011}).
The left-sided Caputo fractional derivative \cite{KST}
of order $\alpha >0$ for $x \in [a,b]$ is defined by
\be \label{4-1}
\,  ^CD^{\alpha}_{a+} f(t)= \, I^{n-\alpha}_{a+} D^n_x f(x) =
\frac{1}{\Gamma(n-\alpha)} \int^x_a 
\frac{ dz \, D^n_{z}f(z)}{(x-z)^{\alpha-n+1}} ,
\ee
where $n-1 < \alpha <n$, and $I^{\alpha}_{a+}$ is 
the left-sided Riemann-Liouville fractional integral 
of order $\alpha >0$ that is defined by
\[ I^{\alpha}_{a+} f(x)=\frac{1}{\Gamma(\alpha)} 
\int^x_a \frac{f(z) d z}{(x-z)^{1-\alpha}} , \quad (z>a). \]

Then using the usual definition of the strain $\varepsilon(x)$
in terms of the displacement $u(x)$ 
\be \varepsilon (x) =D^1_x u(x) , \ee
we obtain the fractional stress-displacement equation in the form
\be \label{s4}
\sigma(x) = E \, D^1_x u (x) \pm l^2_{\beta} \, E \ ^CD^{\beta+1}_{a+} u (x) .
\ee

In view of the fractional vector calculus framework, we can derive 
the fractional equation of equilibrium in the form 
\be \label{3.1fC}
A_{\alpha}(x) \, ^CD^{\alpha}_{a+} \sigma (x) + f(x) =0 
\ee
with a given function $A_{\alpha}(x)$, and $f(x)$ denoting, as usual, the external body force field.
The explicit form of the function $A_{\alpha}(x)$ is derived from 
the conservation law for non-local media by using 
the fractional vector calculus \cite{AP2008}.
Substitution of (\ref{s4}) into (\ref{3.1fC}) gives
\be \label{FE-CD1}
 ^CD^{\alpha+1}_{a+} u(x) \pm l^2_{\beta} \, ^CD^{\alpha}_{a+} \, ^CD^{\beta+1}_{a+} u(x) + \tilde f(x) =0 ,
\ee
where 
\be \label{f} 
\tilde f(x) =E^{-1} A^{-1}_{\alpha}(x) f(x) . 
\ee

For the case $\alpha=1$, the governing fractional differential equation reads
\be \label{FE-CD1b}
D^2_x u (x) \pm l^2_{\beta} \, D^1_x \, ^CD^{\beta+1}_{a+} u (x) + \tilde f(x) =0 .
\ee
In general, we have $D^1_x \, ^CD^{\beta+1}_x \ne \, ^CD^{\beta+2}_x$.

%%%%%%%%%%%%%%%%%%%%%%%%%%%%%%%%%%%%%%%%%%%%%%%%%%%%%%%%%%%%%%%%%%%%%%%%%%%%%%

\subsection{Solution of the fractional gradient elasticity equation}

Let us make use of the explicit form concerning the violation of 
the semigroup property for the Caputo derivative
that gives the relationship between 
the product $\, ^CD^{\alpha}_{a+} \, ^CD^{\beta}_{a+} $ and 
the derivative $\, ^CD^{\alpha+\beta}_{a+}$.

Using Eq. 2.4.6. in \cite{KST} of the form
\be
(\, ^CD^{\alpha}_{a+} f)(x)\, = (\, ^{RL}D^{\alpha}_{a+} f)(x) -
\sum^{n-1}_{k=0} \frac{(D^k f)(a)}{\Gamma(k-\alpha+1)} (x-a)^{k-\alpha} ,
\ee
and using Property 2.1, Eq. 2.1.16 in \cite{KST}, 
\be \label{Int-a}
I^{\alpha}_{a+} (x-a)^{\beta} = 
\frac{\Gamma(\beta+1)}{\Gamma(\alpha+\beta)} (x-a)^{\beta+\alpha} ,
\ee
where $\alpha>0$ and $\beta>-1$, we obtain the relation
\be \label{VSP}
^CD^{\alpha}_{a+} \, ^CD^{\beta}_{a+} f(x) =
\, ^CD^{\alpha+\beta}_{a+} f(x) +
\sum^{n-1}_{k=0} \frac{f^{(k+n)}(a)}{\Gamma(n+k-\alpha-\beta)} 
(x-a)^{n+k-\alpha-\beta} ,
 \ee
where $0<\alpha \le 1$, $n-1<\beta \le n$.
This relation explicitly shows a violation of 
the semigroup property for the Caputo derivative. 

Using (\ref{VSP}), we rewrite (\ref{FE-CD1}) in the form
\be \label{FE-CD4}
(\, ^CD^{\alpha+\beta+1}_{a+} u)(x) \pm 
l^{-2}_{\beta} (\, ^CD^{\alpha+1}_{a+} u)(x) \pm f_{eff} (x) =0 ,
\ee
where $0<\alpha<1$, $1<\beta<2$ ($n=2$), or $2<\beta<3$ ($n=3$), and
$f_{eff} (x)$ is an effective body force defined by
\be
f_{eff} (x) = l^{-2}_{\beta} 
\sum^{n}_{k=0} \frac{u^{(k+n+1)}(a)}{\Gamma(n+k-\alpha-\beta)} 
(x-a)^{n+k-\alpha-\beta} + l^{-2}_{\beta} \, \tilde f(x)  .
\ee
Equation (\ref{FE-CD4}) is a nonhomogeneous fractional differential equation 
with constant coefficients.

The solutions to equations of this type are given by Theorem 5.16 in \cite{KST}  
(see also Theorem 5.13 in \cite{KST} for the homogeneous case, $f_{eff}(x)=0$).
To use these results, we assume that $a=0$.
Let us consider the case $0<\alpha<1$, $1<\beta<2$ 
(i.e. $1<\alpha+1<2=m$, $2<\beta+1<3=n$). 
Then the solution of (\ref{FE-CD4}) has the form
\be \label{SOL}
u(x)= \int^x_0 dz \, f_{eff}(z) \, (x-z)^{\alpha} E_{\alpha-\beta,\alpha+1} [ \mp l^{-2}_{\beta} (x-z)^{\alpha-\beta} ]
+ C_0 u_0(x)+ C_1 u_1(x) + C_2 u_2(x)  ,
\ee
where 
\be \label{u-1b}
u_0(x) = E_{\alpha-\beta,1} [ \mp l^{-2}_{\beta} x^{\alpha-\beta}] 
\pm l^{-2}_{\beta} x^{\alpha -\beta} \, E_{\alpha-\beta,\alpha-\beta+1} [ \mp l^{-2}_{\beta} x^{\alpha-\beta}] ,
\ee
\be \label{u-1c}
u_1(x) = x \, E_{\alpha-\beta,2} [ \mp l^{-2}_{\beta} x^{\alpha-\beta}] 
\pm l^{-2}_{\beta} x^{\alpha -\beta+1} \, E_{\alpha-\beta,\alpha-\beta+2} [ \mp l^{-2}_{\beta} x^{\alpha-\beta}] ,
\ee
\be  \label{u-2b}
u_2(x) =  x^2 E_{\alpha-\beta,3} [ \mp l^{-2}_{\beta} x^{\alpha-\beta}] ,
\ee
and
\be \label{eff}
f_{eff}(x) = l^{-2}_{\beta} 
\sum^{2}_{k=0} \frac{u^{(k+3)}(0)}{\Gamma(2+k-\alpha-\beta)} 
x^{2+k-\alpha-\beta} + l^{-2}_{\beta} \, \tilde f(x)  .
\ee
The arbitrary real constants $C_0$, $C_1$, and $C_2$ 
in the case of the Caputo fractional derivatives 
are defined by the values of 
the integer-order derivatives $u(0)$, $u^{(1)}(0)$, and $u^{(2)}(0)$.

Here $E_{\alpha,\beta}(z)$ is the Mittag-Leffler function \cite{KST} that is defined by
\be \label{Eab}
E_{\alpha,\beta}[z] = \sum^{\infty}_{k=0} \frac{z^k}{\Gamma(\alpha k+\beta)} , 
\quad (\alpha>0, \beta \in \mathbb{R}) .
\ee
Note also that $E_{1,1}[z]=e^z$.
The asymptotic behavior (see Eq. 1.8.27 in \cite{KST}) 
of the Mittag-Leffler function $E_{\alpha,\beta}(z)$ is  
\be
E_{\alpha,\beta}(z) = \frac{1}{\alpha} z^{(1-\beta)/\alpha} \, exp (z^{1/\alpha})
-\sum^N_{k=1} \frac{1}{\Gamma(\beta - \alpha k)} \frac{1}{z^k} + O(1/ z^{N+1})  \quad (|z| \to \infty),
\ee
where $0<\alpha<2$. \\

{\bf Remark:} 
It should be emphasized that the absence of the external force ($f(x)=0$) 
does not imply the vanishing of the effective force $f_{eff}$. 
In general, $f_{eff}(x) \ne 0$ for $f(x)=0$.
Only in the case of commutativity of the Caputo fractional derivatives, i.e. 
if the semigroup property 
\[ (\, ^CD^{\alpha}_{a+} \, ^CD^{\beta}_{a+} u)(x) = (\, ^CD^{\alpha+\beta}_{a+} u)(x) \]
is not violated, the vanishing of the external force $f(x)=0$ 
leads to the vanishing of the effective force $f_{eff}(x)=0$. 
It is easy to see that the semigroup property is satisfied if
\be \label{u345}
u^{(3)}(0) = u^{(4)}(0) = u^{(5)}(0) =0 . 
\ee

If we consider (\ref{FE-CD4}) in the case $\alpha=1$, $1<\beta<2$, $f(x)=0$ and 
assume that condition (\ref{u345}) is satisfied, then the solution (\ref{SOL})
of (\ref{FE-CD4}) has the form
\[ u(x) = C_0 \Bigl( E_{1-\beta,1} [ \mp l^{-2}_{\beta} x^{1-\beta}] 
\pm l^{-2}_{\beta} x^{1 -\beta} \, E_{1-\beta,2-\beta} [ \mp l^{-2}_{\beta} x^{1-\beta}] \Bigr) + \]
\be \label{SS1} 
+ C_1  \Bigl( x \, E_{1-\beta,2} [ \mp l^{-2}_{\beta} x^{1-\beta}] 
\pm l^{-2}_{\beta} x^{2 -\beta} \, E_{1-\beta,3-\beta} [ \mp l^{-2}_{\beta} x^{1-\beta}] \Bigr) +
C_2 x^2 E_{1-\beta,3} [ \mp l^{-2}_{\beta} x^{1-\beta}] .
\ee
In order for this solution to be admissible, 
it should be checked if the function given by (\ref{SS1})
satisfies the conditions $u^{(3)}(0) = u^{(4)}(0) = u^{(5)}(0) =0$.
To verify these conditions we use Eq. (1.8.22) of \cite{KST} in the form
\be
\frac{d^n}{dz^n} E_{\alpha,\beta} [z] = n! \, E^{n+1}_{\alpha,\beta+ n \alpha} [z] ,
\quad (n \in \mathbb{N}) .
\ee
The conditions given by (\ref{u345}) are not satisfied for the 
function given by (\ref{SS1}).
Thus, the solution is not admissible for the fractional one-dimensional 
gradient elasticity model considered herein.
Therefore, we should take into account the effective force 
defined in (\ref{eff}) in order for the solution given by (\ref{SOL}) 
to describe the fractional one-dimensional gradient elasticity model correctly.

%%%%%%%%%%%%%%%%%%%%%%%%%%%%%%%%%%%%%%%%%%%%%%%%%%%%%%%%%%%%%%%%%%%%%%%%%%%%%%%%%%%%%%%%%%%%%%%%%%%%%%%%
%%%%%%%%%%%%%%%%%%%%%%%%%%%%%%%%%%%%%%%%%%%%%%%%%%%%%%%%%%%%%%%%%%%%%%%%%%%%%%%%%%%%%%%%%%%%%%%%%%%%%%%%
%%%%%%%%%%%%%%%%%%%%%%%%%%%%%%%%%%%%%%%%%%%%%%%%%%%%%%%%%%%%%%%%%%%%%%%%%%%%%%%%%%%%%%%%%%%%%%%%%%%%%%%%

\section{Fractional gradient elasticity based on Riesz derivative}

An alternative fractional gradient elasticity model may be obtained by employing
the Riesz fractional derivative. 
In this case, it turns out that a three-dimensional treatment is possible
due to available results on the fractional Laplacian of the Riesz type.
The corresponding fractional gradient elasticity governing equation 
can then be considered in the form
\be \label{FPDE-4}
c_{\alpha} \, ((-\Delta)^{\alpha/2} u) ({\bf r}) + 
c_{\beta} \, ((-\Delta)^{\beta/2} u) ({\bf r}) = f({\bf r})  \quad (\alpha> \beta),
\ee
where ${\bf r} \in \mathbb{R}^3$ and $r=|{\bf r}|$ are dimensionless,
$(-\Delta)^{\alpha /2}$ is the Riesz fractional Laplacian of order $\alpha$ \cite{KST}. 
The coefficients ($c_{\alpha}$, $c_{\beta}$) are phenomenological constants and 
the rest of the symbols have their usual meaning.
For $\alpha > 0$ and suitable functions $u({\bf r})$,
${\bf r} \in \mathbb{R}^3$, the Riesz fractional derivative
can be defined \cite{KST} in terms of the Fourier transform ${\cal F}$ by
\be
((-\Delta)^{\alpha /2} u)({\bf r})= {\cal F}^{-1} \Bigl( |{\bf k}|^{\alpha} ({\cal F} u)({\bf k}) \Bigr) ,
\ee
where ${\bf k}$ denotes the wave vector.
If $\alpha=4$ and $\beta=2$, we have the well-known equation of the gradient elasticity \cite{AA2011}:
\be \label{GradEl}
c_2 \, \Delta u ({\bf r}) - c_4 \Delta^2 u ({\bf r}) + f({\bf r}) = 0,
\ee
where
\be \label{GradEl-2} 
c_2 = E , \quad c_4 = \pm \, l^2 \, E .
\ee

Equation (\ref{FPDE-4}) is the fractional partial differential equation that
has the particular solution (Section 5.5.1. in \cite{KST}) of the form
\be \label{phi-G4}
u({\bf r})= \int_{\mathbb{R}^3} 
G^3_{\alpha, \beta} ({\bf r} - {\bf r}^{\prime}) \, 
f({\bf r}^{\prime}) \, d^3 {\bf r}^{\prime},
\ee
where the Green-type function 
\be \label{FGF}
G^3_{\alpha}({\bf r})= 
\int_{\mathbb{R}^3} \frac{1}{c_{\alpha} |{\bf k}|^{\alpha} + c_{\beta} |{\bf k}|^{\beta}  } \
e^{ + i ({\bf k},{\bf r}) } \, d^3 {\bf k} 
\ee
is given (see Lemma 25.1 of \cite{SKM}) by the equation
\be \label{G-4}
G^3_{\alpha, \beta} ({\bf r}) =\frac{1}{(2 \pi)^{3/2} \, \sqrt{|{\bf r}|}} 
\int^{\infty}_0 \frac{ \lambda^{3/2} \, J_{1/2} (\lambda |{\bf r}|) 
}{c_{\alpha} \lambda^{\alpha}+ c_{\beta} |\lambda|^{\beta}} 
\, d \lambda .
\ee
Here $J_{1/2}(z) = \sqrt{2/(\pi z)} \, \sin (z)$ is the Bessel function of the first kind.

%%%%%%%%%%%%%%%%%%%%%%%%%%%%%%%%%%%%%%%%%%%%%%%%%%%%%%%%%%%%%%%%%%%%%%%%%%%%%%

Let us consider, as an example, the W. Thomson (1848) problem \cite{LL} 
for the present model of fractional gradient elasticity.
Determine the deformation of an infinite elastic continuum,
when a concentrated force is applied to a small region of it.
To solve this problem, we consider distances $|{\bf r}|$, 
which are large in comparison with the size of the region
(neighborhood) of load application. In other words,
we can suppose that the force is applied at a point. 
In this case, we have
\be \label{deltaf}
f({\bf r}) = f_0 \, \delta ({\bf r}) = f_0 \, \delta (x) \delta (y) \delta (z)  . 
\ee
Then, the displacement field $u ({\bf r})$ of fractional gradient elasticity 
has a simple form given by the particular solution 
\be \label{phi-Gb}
u ({\bf r}) = f_0 \, G^3_{\alpha} ({\bf r}) ,
\ee
where $G^3_{\alpha}(z)$ is the Green's function given by (\ref{G-4}). 
Therefore, the displacement field for the force 
that is applied at a point (\ref{deltaf}) has the form
\be \label{Pot-2}
u ({\bf r}) = \frac{1}{2 \pi^2} \frac{f_0}{|{\bf r}|} \, 
\int^{\infty}_0 \frac{ \lambda \, \sin (\lambda |{\bf r}|)}{ 
c_{\alpha} \lambda^{\alpha}+ c_{\beta} \lambda^{\beta}  } \, d \lambda \quad (\alpha > \beta).
\ee

For this solution of the fractional gradient elasticity equation (\ref{FPDE-4}) 
with $\alpha>\beta$, $0<\beta<2$, and $\alpha \ne 2$, 
with a point force $f({\bf r})$ of the form given by  Eq. (\ref{deltaf}),
the asymptotic behavior is
\be
u ({\bf r}) \ \approx \  \frac{f_0 \, \Gamma(2-\beta) \sin(\pi \beta/2)}{2 \pi^2 \, \, c_{\beta}} \,
\cdot \,  \frac{1}{|{\bf r}|^{3-\beta}} \quad (|{\bf r}| \to \infty) .
\ee
This asymptotic behavior $|{\bf r}| \to \infty$ does not depend on the parameter $\alpha$,
and the corresponding asymptotic behavior for $|{\bf r}| \to 0$ 
does not depend on the parameter $\beta$, where $\alpha>\beta$.
The displacement field at large distances from the point of load application 
is determined only by term $(-\Delta)^{\beta/2}$, where $\beta < \alpha$. 
This can be interpreted as a fractional non-local "deformation" 
counterpart of the classical result based on the local Hooke's law. 
We note the existence of a maximum for the quantity 
$u ({\bf r}) \cdot |{\bf r}|$ in the case $0<\beta < 2 < \alpha$. 

From a mathematical point of view, there are two special cases:
(I) fractional power-law weak non-locality with $\alpha=2$ and $0<\beta<2$;
(II) fractional power-law weak non-locality with $\alpha \ne 2$, $\alpha > \beta$ and $0<\beta <3$.
It is thus useful to distinguish between the following two particular cases:

\begin{itemize}

\item
Sub-gradient elasticity ($\alpha=2$ and $0<\beta<2$). 

\item
Super-gradient elasticity ($\alpha > 2$ and $\beta=2$). 

\end{itemize}

In the sub-gradient elasticity model the order of the fractional derivative 
is less than the order of the term related to the usual Hooke's law. 
The order of the fractional derivative in 
the super-gradient elasticity equation is larger than 
the order of the term related to the Hooke's law.
The terms "sub-" and "super-" gradient elasticity are used in analogy 
to the terms commonly used for anomalous diffusion \cite{MK1,Zaslavsky1,MK2}: 
sub-diffusion and super-diffusion. \\

%%%%%%%%%%%%%%%%%%%%%%%%%%%%%%%%%%%%%%%%%%%%%%%%%%%%%%%%%%%%%%%%%%%%%%%%%%%%%%

{\bf (I) Sub-gradient elasticity model}: The fractional model of sub-gradient elasticity 
is described by (\ref{FPDE-4}) with $\alpha=2$ and $0<\beta<2$, i.e. 
\be \label{FPDE-4-2b1}
c_2 \Delta u ({\bf r}) - c_{\beta} ((-\Delta)^{\beta/2} u) ({\bf r}) +  f({\bf r}) = 0 ,
\quad (0<\beta<2) .
\ee
The order of the fractional Laplacian $(-\Delta)^{\beta/2}$ is less 
than the order of the first term related to the usual Hooke's law. 
As a simple example, we can consider the square of the Laplacian, i.e. $\beta=1$.
In general, the parameter $\beta$ defines the order 
of the power-law non-locality that characterizes the elastic continuum.
The particular solution of (\ref{FPDE-4-2b1}) for the point force 
problem at hand, reads
\be \label{Pot-2-2b}
u ({\bf r}) = \frac{1}{2 \pi^2} \frac{f_0}{|{\bf r}|} \, 
\int^{\infty}_0 \frac{ \lambda \, \sin (\lambda |{\bf r}|)}{ 
c_2 \lambda^2+ c_{\beta} \lambda^{\beta}  } \, d \lambda \quad (0< \beta <2).
\ee
The following asymptotic behavior for (\ref{Pot-2-2b})   
can be derived by using Section 2.3.1 in \cite{BE} of the form 
\be
u ({\bf r}) = \frac{f_0}{2\pi^2 \, |{\bf r}|} \,
\int^{\infty}_0 \frac{\lambda \sin (\lambda |{\bf r}|)}{c_2 \lambda^2+ c_{\beta} \lambda^{\beta}}  \, d \lambda \approx
 \frac{ C_0(\beta) }{|{\bf r}|^{3-\beta}} + \sum^{\infty}_{k=1}  \frac{ C_k(\beta) }{|{\bf r}|^{(2-\beta)(k+1)+1}} 
\quad (|{\bf r}| \to \infty) ,
\ee
where 
\be
C_0(\beta)= \frac{f_0}{2\pi^2 \, c_{\beta}} \, \Gamma(2-\beta) \, \sin \left( \frac{\pi}{2}\beta \right) ,
\ee
\be
C_k (\beta) = -\frac{f_0 c^k_2}{2 \pi^2 \, c^{k+1}_{\beta}} \int^{\infty}_0  z^{(2-\beta)(k+1)-1} \, \sin(z) \, dz .
\ee
As a result, the displacement field generated by the force that is applied at a point
in the elastic continuum with the fractional non-locality described 
by the fractional Laplacian $(-\Delta)^{\beta/2}$ with $ 0<\beta <2$
is given by
\be
u ({\bf r}) \ \approx \ \, \frac{C_0 (\beta)}{|{\bf r}|^{3-\beta}} \quad (0< \beta<2),
\ee
for large distances $|{\bf r}| \gg 1$.  \\

%%%%%%%%%%%%%%%%%%%%%%%%%%%%%%%%%%%%%%%%%%%%%%%%%%%%%%%%%%%%%%%%%%%%%%%%%%%%%%

{\bf (II) Super-gradient elasticity model}: 
The fractional model of super-gradient elasticity 
is described by (\ref{FPDE-4}) with $\alpha > 2$ and $\beta=2$.
In this case we have 
\be \label{FPDE-4-2b2}
c_2 \Delta u ({\bf r}) - c_{\alpha} ((-\Delta)^{\alpha/2} u) ({\bf r}) + f({\bf r}) = 0 ,
\quad (\alpha > 2) .
\ee
The order of the fractional Laplacian $(-\Delta)^{\alpha/2}$ 
is greater than the order of the first term related to the usual Hooke's law.
The parameter $\alpha >2$ defines the order of the power-law non-locality of the elastic continuum.
If $\alpha =4$, (\ref{FPDE-4-2b2}) reduced to (\ref{GradEl}). 
The case $3<\alpha<5$ can be considered to correspond
as closely as possible ($\alpha \ \approx \ 4$)
to the usual gradient elasticity model of Eq. (\ref{GradEl}).

The asymptotic behavior of the displacement field $u (|{\bf r}|)$ for $|{\bf r}| \to 0$ 
in the case of super-gradient elasticity is given by
\be \label{Cab-2}
u ({\bf r}) \ \approx \ 
\frac{f_0  \, \Gamma((3-\alpha)/2)}{2^{\alpha} \, \pi^2 \sqrt{\pi} \, c_{\alpha} \, \Gamma(\alpha/2)} \,
\cdot \,  \frac{1}{|{\bf r}|^{3-\alpha}} , \quad (2<\alpha<3),
\ee
\be \label{Cab-3}
u ({\bf r}) \ \approx \ 
\frac{f_0}{2 \pi \, \alpha  \, c^{1-3/\alpha}_{\beta} \, c^{3/\alpha}_{\alpha} \, \sin (3 \pi / \alpha)}
 , \quad (\alpha>3) .
\ee
Note that the above asymptotic behavior does not depend on the parameter $\beta$, 
and that the corresponding relation (\ref{Cab-2}) does not depend on $c_{\beta}$.
The displacement field $u ({\bf r})$ for short distances away from the point of load
application is determined only by the term with $(-\Delta)^{\alpha/2}$ ($\alpha>\beta$) 
which can be considered as a fractional counterpart of the usual extra non-Hookean term 
of gradient elasticity. 

%%%The idea of a fractional generalization of the phenomenological theory of gradient elasticity 
%%%by including spatial fractional derivatives was suggested by the second author,
%%%and its realization with the Caputo and Riesz fractional derivatives has
%%%been performed by the first author.

A generalization of the phenomenological theory of gradient elasticity 
by including Caputo and Riesz spatial derivatives of non-integer order 
is suggested in the present paper.
Related lattice models with spatial dispersion of power-law type as microscopic models 
of the fractional elastic continuum described by (\ref{FPDE-4}) 
was proposed 
%%%by the first author 
in \cite{arXiv2013-1}.
Using the approach suggested in \cite{JPA2006,JMP2006}, 
equations (\ref{FPDE-4}) 
%%%phenomenological fractional elasticity model 
has been derived from 
the equations of lattice dynamics with power-law spatial dispersion. 
We can point out that a phenomenological fractional gradient elasticity model 
can be obtained from different microscopic or lattice models. 
In addition, we note that the model of fractional gradient elastic continuum 
has an analog in the plasma-like dielectric material with power-law spatial dispersion \cite{AP2013}.
It can be considered as a common or universal behavior of plasma-like and elastic materials 
in space by analogy with the universal behavior of low-loss dielectrics in time \cite{Jo1,Jo2,JPCM2008}.

%%%%%%%%%%%%%%%%%%%%%%%%%%%%%%%%%%%%%%%%%%%%%%%%%%%%%%%%%%%%%%%%%%%%%%%%%%%%%%%%%%%%%%%%%

\section*{Acknowledgments}

The first author extends his thanks to Professor Juan J. Trujillo 
for valuable discussions of applications fractional models in elasticity theory, 
and to Aristotle University of Thessaloniki for its support and kind hospitality in July 2013.

The financial support of the ERC-13 (IL-GradMech-AMS) grant through the Greek-Secretariat
of Research and Technology is gratefully acknowledged.

%%%%%%%%%%%%%%%%%%%%%%%%%%%%%%%%%%%%%%%%%%%%%%%%%%%%%%%%%%%%%%%%%%%%%%%%%%%%%%%%%%%%%%%%%%%%%%%%%%%%%%%%
%%%%%%%%%%%%%%%%%%%%%%%%%%%%%%%%%%%%%%%%%%%%%%%%%%%%%%%%%%%%%%%%%%%%%%%%%%%%%%%%%%%%%%%%%%%%%%%%%%%%%%%%
%%%%%%%%%%%%%%%%%%%%%%%%%%%%%%%%%%%%%%%%%%%%%%%%%%%%%%%%%%%%%%%%%%%%%%%%%%%%%%%%%%%%%%%%%%%%%%%%%%%%%%%%

%%%%%%%%%%%%%%%%%%%%%%%%%%%%%%%%%%%%%%%%%%%%%%%%%%%%%%%%%%%%%%%%%%%%%%%%%%%%%%%%%%%%%%%%%%%%%%%%%%%%%%%%

\end{document}